\documentclass[aps,prl,twocolumn,amsmath,amssymb, superscriptaddress]{revtex4-2}
\usepackage{graphicx}
\usepackage{subfigure}
\usepackage{epsfig}
\usepackage{ulem}
\usepackage{dcolumn}
\usepackage{bm}
\usepackage{hyperref}
\hypersetup{colorlinks=true, citecolor=blue, urlcolor=blue, linkcolor=blue}
\renewcommand{\v}[1]{{\boldsymbol{#1}}}

\newcommand{\imth}{\hspace{1pt}\mathrm{i}\hspace{1pt}}
\newcommand{\dif}{\mathrm{d}}

\newcommand{\E}[1]{{\mathrm{e}^{ #1}}}

\newcommand{\ra}{\rightarrow}
\newcommand{\s}{{\sigma}}

\def\beq{\begin{equation}}
\def\eeq{\end{equation}}
\def\bald{\begin{aligned}}
\def\eald{\end{aligned}}
\def\bea{\begin{eqnarray}}
\def\eea{\end{eqnarray}}

\def\inc#1{\left(#1\right)}

\def\ket#1{\left|#1\right\rangle}
\def\avg#1{\left\langle#1\right\rangle}
\def\braket#1#2{\left\langle #1\right|\left.#2\right\rangle}
\def\expectation#1#2#3{\left\langle #1\left| #2 \right| #3\right\rangle}

\def\Eq#1{Eq.~(\ref{#1})}
\def\Fig#1{Fig.~\ref{#1}}

\begin{document}
\title{Robustness of Antiferromagnetism in the Su-Schrieffer-Heeger-Hubbard model}
\author{Xun Cai}
\affiliation{Beijing National Laboratory for Condensed Matter Physics and Institute of Physics, Chinese Academy of Sciences, Beijing 100190, China}
\affiliation{Institute for Advanced Study, Tsinghua University, Beijing, 100084, China.}
\author{Zi-Xiang Li}
\email{zixiangli@iphy.ac.cn}
\affiliation{Beijing National Laboratory for Condensed Matter Physics and Institute of Physics, Chinese Academy of Sciences, Beijing 100190, China}
\author{Hong Yao}
\email{yaohong@tsinghua.edu.cn}
\affiliation{Institute for Advanced Study, Tsinghua University, Beijing, 100084, China.}
\affiliation{State Key Laboratory of Low Dimensional Quantum Physics, Tsinghua University, Beijing 100084, China.}

\begin{abstract}
We recently unveiled that antiferromagnetism (AFM) order can be dominantly induced by the bond Su-Schrieffer-Heeger (SSH) electron-phonon coupling (EPC), which is beyond the conventional wisdom that AFM order is usually driven by strong Coulomb interactions. Nevertheless, many aspects of the interplay between EPC and strong electronic interactions on AFM ordering remains unexplored. Here, we investigate the Su-Schrieffer-Heeger-Hubbard (SSHH) model with bond SSH phonons and onsite Hubbard interactions by large-scale quantum Monte-Carlo (QMC) simulations and obtain its ground-state phase diagram for various Hubbard interactions and EPC strength at half filling. Our results show that Hubbard interactions further enhance EPC-induced AFM, especially for small phonon frequency or in adiabatic limit, the regime most relevant to various quantum materials. 
This could shed further light to understanding the cooperative effect between EPC and electronic correlations in quantum materials.
\end{abstract}
\date{\today}

\maketitle
{\bf Introduction:} Electron-phonon coupling (EPC) and electronic Coulomb interaction are two essential ingredients contributing to many important physical properties of quantum materials. 
It has been shown that EPC is vital in driving charge-density wave (CDW) \cite{Peierls1955book, Grunner1988RMP}, certain topological phases \cite{originalSSH, SSH-review},
and, notably, conventional BCS superconductivity (SC) \cite{BCSoriginal, Schrieffer1964book}. Nonetheless, antiferromagnetism (AFM) such as Neel ordering is widely believed to be triggered mainly by strong Coulomb interactions. Intriguingly, in a recent work of us \cite{XC2021PRL}, 
we convincingly showed for the first time that AFM can be dominantly induced by the EPC between electrons and bond Su-Schrieffer-Heeger (SSH) phonons.

In many quantum materials, both EPC and strong electronic interactions are present; under certain circumstances they can cooperatively affect physical properties of such systems. Indeed, in the past many years, increasing experimental and theoretical progresses suggest that EPC can play an important role even in quantum systems with strong electronic correlation, for instance, in cuprates \cite{Lanzara2001Cuprate,ZXShen2002PhilosophicalReview, ZXShen2004PRL, Nagaosa2004PRL, ZXShen2005PRL-phonon-cuprate, Nagaosa2005ARPESreview, ZXShen2005PRL, Cooper2005PRL, Davis2006Nature-phonon-no-effect, DevereauxCuprate2010PRB, ZXShen2017Science, ZXShenARPES,ZaanenReviewCuprate,Zhang2016Phonon,Xue2016cuprate, Xue2021, Kivelson2019ThermalHall,PhysRevLett.105.257001,PhysRevB.98.035102,chen2021anomalously,PhysRevLett.127.197003,qu2021spin,He2021RMP,he2021superconducting}, iron-based superconductors \cite{Wang2012CPL,ZXShen2014Nature-replicabandFeSe, ZXLi2016FeSe, Johnston2016PRB,DLFeng2017FeSe, JDGuo2017FeSe,Millis2017FeSe,Moses2018SciAdv-FeSe, ZXLi2019PRB-smallmomentum, RuiPeng2020SciAdv,Hoffman2017FeSeReview,DHLee2018Review}, and in various model systems \cite{ZYHan2020PRL,huang2021pair,Weber2021,costa2020phase,sous2017phonon}.  
Various previous works mainly focused on the competition between AFM order induced by Coulomb interactions and other types of orders such as EPC-induced CDW \cite{Assaad1996PRL-dwave, Scalapino1997PRB, Sawatzky2004EPC, Millis2007EPC, Johnston2013PRB, Imada2017EPC, Serella2019arXiv, Campbell2003PRB, Costa2020arx, Hohenadler2019PRB, Kivelson2020nematic, DHLee2007PRB-fRG-dwave, Hague2007Bipolaron, Wang2015Holstein, Wang2020EPC, ShiweiZhang2020PRB,Assaad2018,Anderson1987RVB,Kivelson2003RMP, Anderson2004-vanillaRVB, Wen2006RMP,Scalapino2012RMP,DHLee2013AFM}. 
Since AFM can be induced by EPC with bond SSH phonons \cite{XC2021PRL,goetz2021langevin} or Hubbard interactions \cite{Scalapino2012RMP}, it is desired to understand the effect of Coulomb interactions on the EPC-induced AFM when both interactions are present, especially for small phonon frequency or in adiabatic limit, the regime mostly relevant to various quantum materials. 

\begin{figure}[t]
    \includegraphics[width=0.68\linewidth]{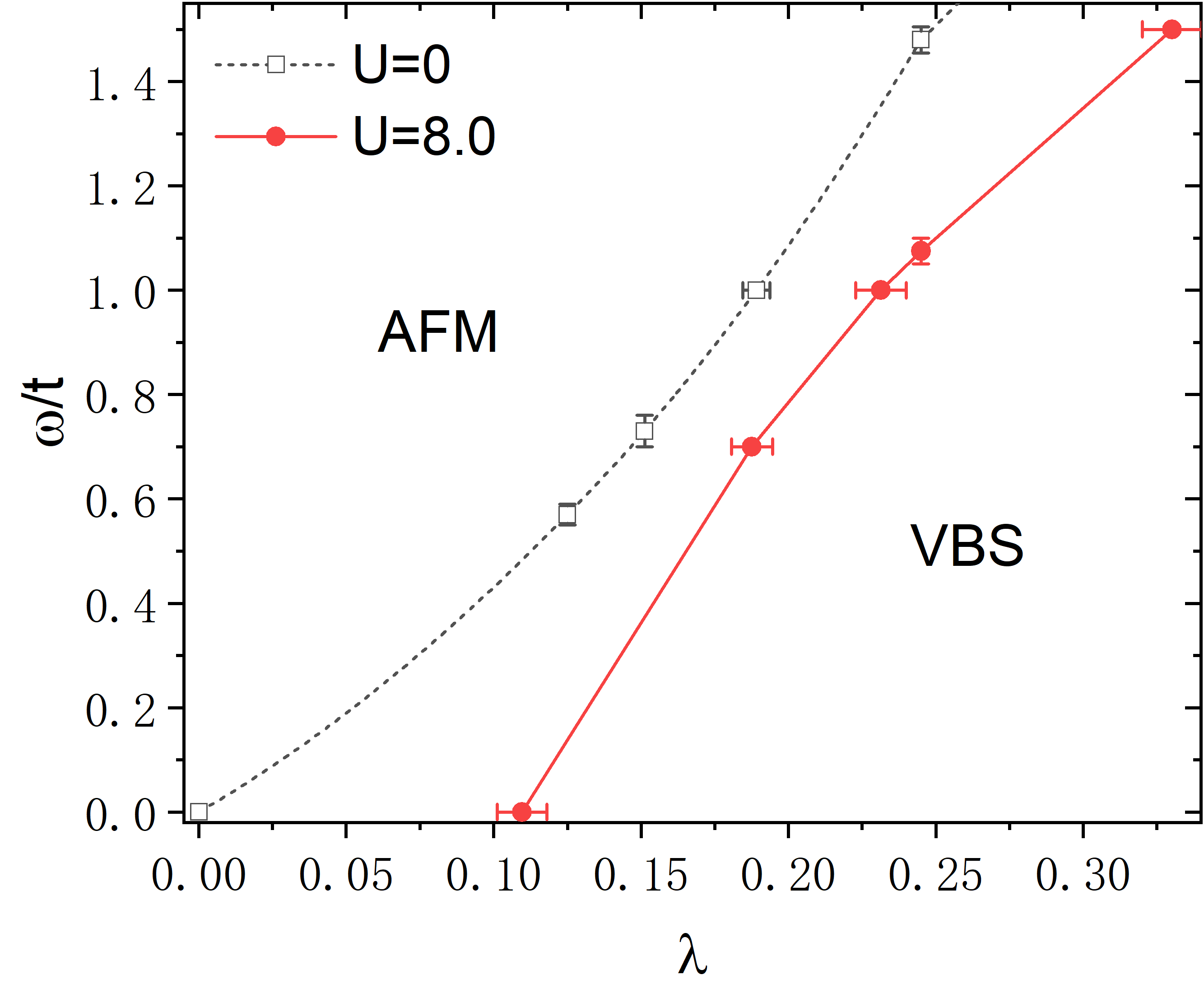}
    \caption{The quantum phase diagram of the half-filled SSHH model as a function of dimensionless EPC constant $\lambda$ and phonon frequency $\omega$ for $U=8t$. The results are obtained by large-scale sign-problem-free projector QMC simulations. The black dashed line indicates the phase boundary of the pure SSH model (namely $U=0$) obtained in Ref. \cite{XC2021PRL}.}
    \label{FigPhaseDiagramU8}
\end{figure}

In the present paper, we investigate the effect of electronic interaction on the EPC-induced AFM ordering by systematically studying the quantum lattice model featuring both EPC of bond SSH phonons \cite{XC2021PRL,goetz2021langevin,Scalettar2021,Sous2018,Sous2021,Assaad2015SSH, Hohenadler2020SSH} and Hubbard electronic interactions, dubbed as Su-Schrieffer-Heeger-Hubbard (SSHH) model. To the best of our knowledge, the bond SSHH model was first introduced in Ref.~\cite{XC2021PRL}. Importantly, in Ref. \cite{XC2021PRL} we noticed that the SSHH model at half-filling is sign-problem-free  \cite{ZXLi2015PRB,ZXLi2016PRL,TXiang2016PRL,Berg2012Science,CJWu2005PRB, Troyer2005PRL,LWang2015PRL} (for a recent review on sign-problem-free QMC, see Ref. \cite{ZXLiQMCreview}) and we performed large-scale QMC simulation \cite{BSS1981PRD,Assaadnote} to access the ground-state properties of the model with large system sizes. We showed that \cite{XC2021PRL}, with vanishing or weak Hubbard interactions, AFM ordering can be induced by the SSH phonons in a large parameter regime of phonon frequency and EPC strength. Here, to reveal the effect of strong electronic correlation on the SSH-phonon induced AFM, we consider strong Hubbard interactions in the simulation of the SSHH model, and obtain its ground-state phase diagram of the SSHH model by varying phonon frequency $\omega$ and EPC strength $\lambda$, as shown in \Fig{FigPhaseDiagramU8}. The obtained phase diagram for $U=8t$ reveals that the Hubbard interactions can further enhance EPC-induced AFM. The enhancement is mostly pronounced in the regime where the SSH phonon frequency is small, which is practically relevant in various realistic quantum materials. We notice a recent parallel work to the one presented here also studies the interesting interplay between EPC of bond SSH phonons and Hubbard interactions although it focuses on different physical aspects and different parameter regime \cite{Richard21}.

{\bf Model:} We consider the SSHH model with bond SSH phonons and onsite Hubbard interactions on the square lattice at half filling, which was initially studied in Ref. \cite{XC2021PRL}. It is described by the following Hamiltonian
\bea\label{EqOriginalModel}
&&H=-\sum_{\avg{ij}}(t-g\hat{X}_{ij})(c^\dagger_{i\s}c_{j\s}+h.c.) +\sum_{\avg{ij}}\frac{\hat{P}^2_{ij}}{2M}+\frac{K}{2}\hat{X}^2_{ij} \nonumber\\
&&~~~~~+U\sum_i(n_{i\uparrow}-\frac{1}{2})(n_{i\downarrow}-\frac{1}{2}),
\eea
where $c^\dagger_{i\s}$ creates an electron on site $i$ with spin polarization $\s=\uparrow$/$\downarrow$, $\avg{ij}$ refers to lattice bonds between nearest neighbor (NN) sites. $\hat{X}_{ij}$ and $\hat P_{ij}$ are the displacement and momentum operators of the optical SSH phonons on each NN bond $\avg{ij}$, with phonon frequency $\omega=\sqrt{K/M}$. Here $t$ is the electron's NN hopping amplitude. The chemical potential $\mu$ is implicit in the Hamiltonian and we focus on the case of half-filling by setting $\mu=0$. Hereafter we set $t=1$ as energy unit and $K=1$ by appropriately redefining displacement fields $\hat{X}_{ij}$. The SSH-type phonons feature linear coupling to electron's NN hopping rather than electron density. To characterize the EPC strength, we define a dimensionless constant $\lambda\equiv\frac{g^2/K}{W}$, where $W=8t$ is the characteristic band width of the square lattice. The last term in Hamiltonian \Eq{EqOriginalModel} denotes on-site electronic Hubbard repulsion with $U>0$.

The pure SSH model without Hubbard interactions is systematically studied in recent works \cite{XC2021PRL,goetz2021langevin,Scalettar2021}. At half filling, the model respects both spin and pseudo-spin SU(2) symmetries \cite{SCZhang1990PRL}. 
The ground state of the pure SSH model was shown to be degenerate between the state with spin ordering and the state with pseudo-spin AFM orderings for a large region of EPC constant $\lambda$ and phonon frequency $\omega$. This degeneracy at $U=0$ originates from the additional particle-hole $\mathbb{Z}_2$ symmetry for spin-down (or spin-up) electrons,  $c_{i\downarrow}\to (-1)^i c^\dag_{i\downarrow}$, in the pure SSH model, which transforms the spin AFM ordering into the pseudo-spin AFM ordering or vice versa, giving rise to the degeneracy between states with spin ordering or pseudo-spin AFM ordering \cite{XC2021PRL}. While for large $\lambda$ and small $\omega$, staggered valence bond solid (VBS) ordering prevails over AFM ordering in the SSH model. In the presence of Hubbard interaction ($U>0$), the spin-down particle-hole $\mathbb{Z}_2$ symmetry breaks down and spin AFM ordering tendency becomes dominant over pseudo-spin AFM ordering. Hereafter, we focus on discussing the tendency of spin AFM and staggered VBS orderings in the SSHH model, and investigating the interplay between Hubbard interaction and EPC on AFM ordering. Details for the definition of various order parameters are shown in Supplemental Material.

At half-filling, the SSHH model in \Eq{EqOriginalModel} is free from the notorious sign problem so that QMC simulation is available to investigate the ground-state properties with large system size. We perform large-scale projector (namely zero-temperature) QMC algorithm to simulate the zero-temperature properties of the model directly, with details of the algorithm shown in the Supplemental Material. To characterize the possible long-range ordering at zero temperature,  we compute the structure factor $S(\v{q},L)\!=\!\frac{1}{N^2}\sum_{i,j}\text{e}^{\text{i}\v{q}\cdot (\v{r_i}-\v{r_j})}\langle \hat O_i\hat O_j\rangle$ of the corresponding order $\hat{O}$ and evaluate the correlation ratio of the structure factor $R^S(L)\!=\!1-\frac{S(\v{Q}+\v{\delta q},L)}{S(\v{Q},L)}$, where $\v{Q}$ refers to the ordering momentum and $\v{\delta q}\!=\!(\frac{2\pi}{L},\frac{2\pi}{L})$ denotes the minimal momentum shift from $\v{Q}$ on a lattice with size $L\times L$. In the thermodynamic limit ($L\ra\infty$), an ordered phase of $\hat{O}$ features $R^S\ra 1$ while a disordered phase features $R^S\ra 0$, and the correlation ratio is RG-invariant, making it possible to identify the transition point by performing finite size scaling on $R^S$. Additionally, we also compute the susceptibility and the corresponding RG-invariant ratio for AFM ordering $R^\chi(L)\!=\!1-\frac{\chi\inc{\v{Q}+\v{\delta q},L}}{\chi\inc{\v{Q},L}}$, which has smaller finite-size corrections than the correlation ratio \cite{AssaadPRB91_165108}. Details for evaluating the magnetic susceptibilities and the corresponding results can be found in Supplemental Material.

{\bf Results in adiabatic limit (zero frequency):}
Effective interactions between electrons mediated by phonons with a finite frequency are retarded. The ratio between phonon frequency $\omega$ and Fermi energy of electrons is the key quantity to determine the magnitude of retardation. It is widely known that retardation of phonon mediated electronic interactions is responsible for many novel quantum physics, especially superconductivity. In our recent work, it was shown that retardation effect can play crucial roles in SSH phonon driven AFM ordering \cite{XC2021PRL}. Specifically, in the pure SSH model with $U=0$, a direct phase transition from VBS to AFM phase occurs by increasing phonon frequency. Thus, it is intriguing to investigate the effect of strong electronic correlation in the parameter regimes with different phonon frequency, or equivalently, magnitude of retardation. In realistic quantum materials, the phonon frequency is commonly much smaller than electronic Fermi energy, namely, in the adiabatic regime $\omega \ll E_f $. Hence, for simplicity, before discussing the general case at finite phonon frequency, we investigate the SSHH model and explore the effect of electronic correlation on the AFM ordering at the adiabatic limit $\omega=0$, which is closely relevant to realistic quantum materials and expected to deepen our understanding of AFM ordering in correlated quantum materials with both strong electronic interactions and EPC.

In the adiabatic limit $\omega=0$, the kinetic term of phonon in \Eq{EqOriginalModel} vanishes, namely the phonon is static at zero temperature. Under this condition, for $U=0$ the exact solution is available by treating the phonon displacement configuration $X_{ij}$ as variational parameters. Owing to the absence of quantum fluctuation, AFM ordering cannot be induced by the EPC at zero frequency. The exact solution shows that the staggered VBS ordering with momentum $\v{Q}=\inc{\pi,\pi}$, which breaks both lattice translational symmetry and $\mathbb{C}_4$ rotational symmetry, is observed for any finite EPC constant $\lambda$ \cite{XC2021PRL}.

\begin{figure}[t]
\subfigure{\label{figAFMratioU8w0}\includegraphics[width= 0.475\linewidth]{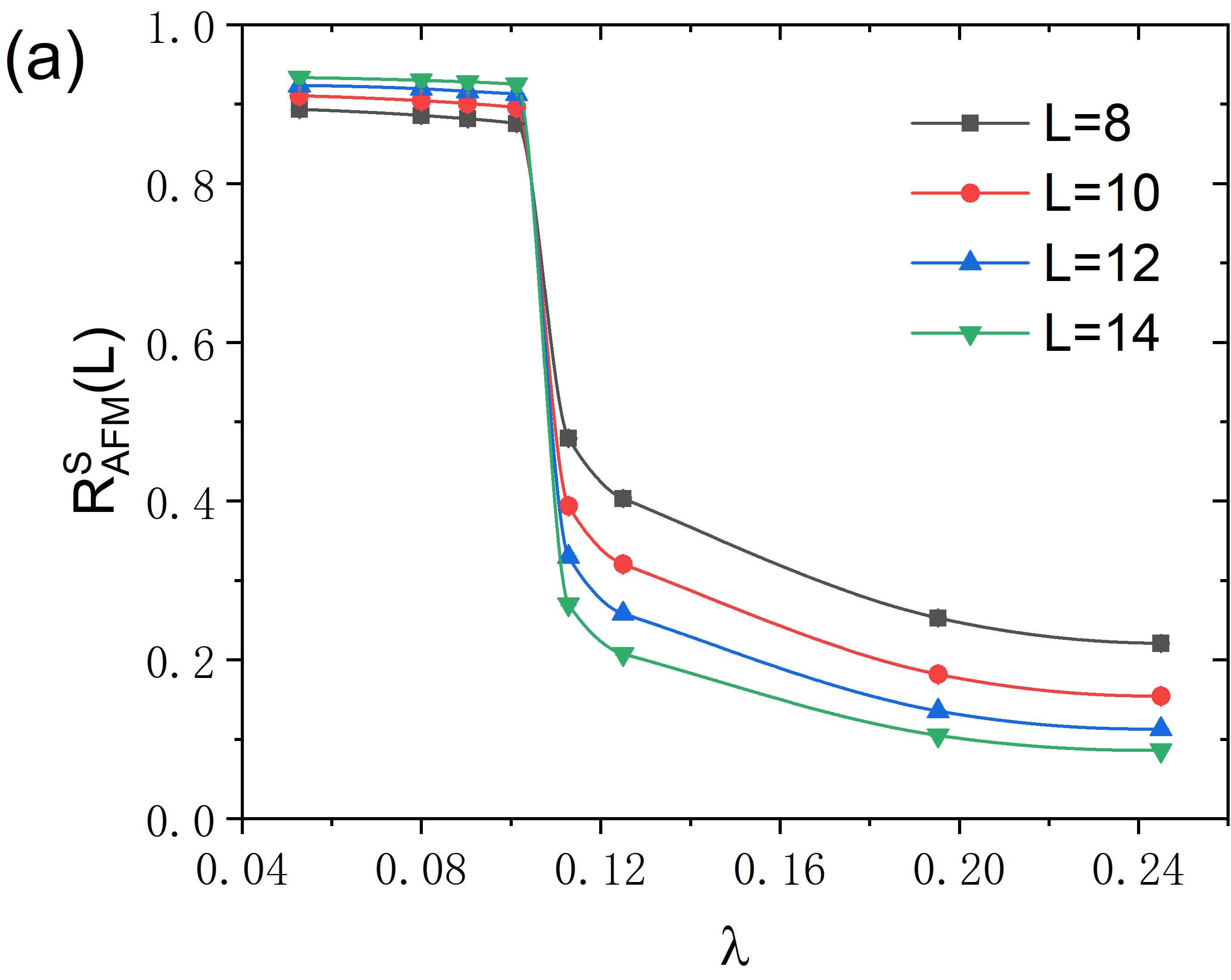}}~~~~
\subfigure{\label{figVBSratioU8w0}\includegraphics[width= 0.475\linewidth]{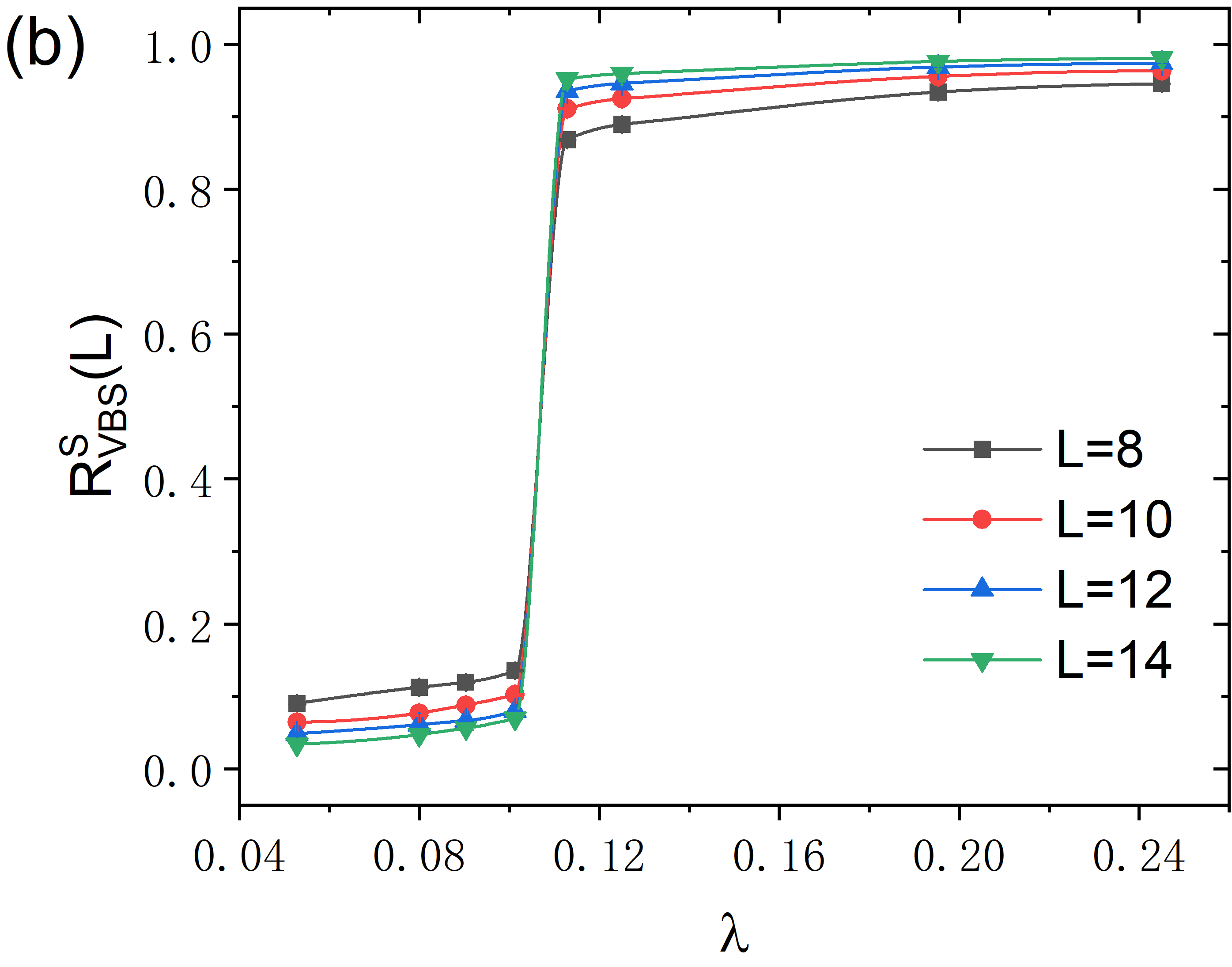}}
\caption{The QMC results of correlation ratios as a function of $\lambda$ with fixed $U=8$ for AFM ordering (a) and VBS ordering (b). The crossing points of correlation ratios for different $L$ suggest that the transition points of AFM and VBS coincide at $\lambda_c\approx 0.11$. }
\label{FigRatioU8w0}
\end{figure}

Nevertheless, the on-site Hubbard repulsion $U$ introduces quantum fluctuations into the system, giving rise to AFM instability due to the super-exchange interactions induced by strong Hubbard interactions. Therefore, with increasing Hubbard repulsion $U$, we expect that AFM ordering tendency will be enhanced and compete with staggered VBS ordering. By employing projector QMC, we systematically explore the effect of electronic correlations in the SSHH model in the adiabatic limit. We evaluate the VBS and AFM correlation ratio as a function of $\lambda$, as shown in \Fig{FigRatioU8w0}. For fixed $U=8t$, the AFM ordered phase can persist up to a critical value of $\lambda_c$, while VBS order appears at larger $\lambda$. The crossing of AFM correlation ratio $R_{\mathrm{AFM}}^S$ for different system sizes approximately coincides with that of VBS correlation ratio $R_{\mathrm{VBS}}^S$ at $\lambda_c\approx 0.11$, indicating the occurrence of a direct AFM-VBS transition. The existence of the finite critical value $\lambda_c$ is qualitatively distinct to the pure SSH model\cite{XC2021PRL} at the adiabatic limit, where VBS ordering appears for any EPC strength. Thus, we conclude that AFM ordering is dominantly introduced by the electronic correlation in the limit of zero frequency (the adiabatic limit) and it can persist up to a sizeable EPC strength.

\begin{figure}[t]
    \includegraphics[width=0.6\linewidth]{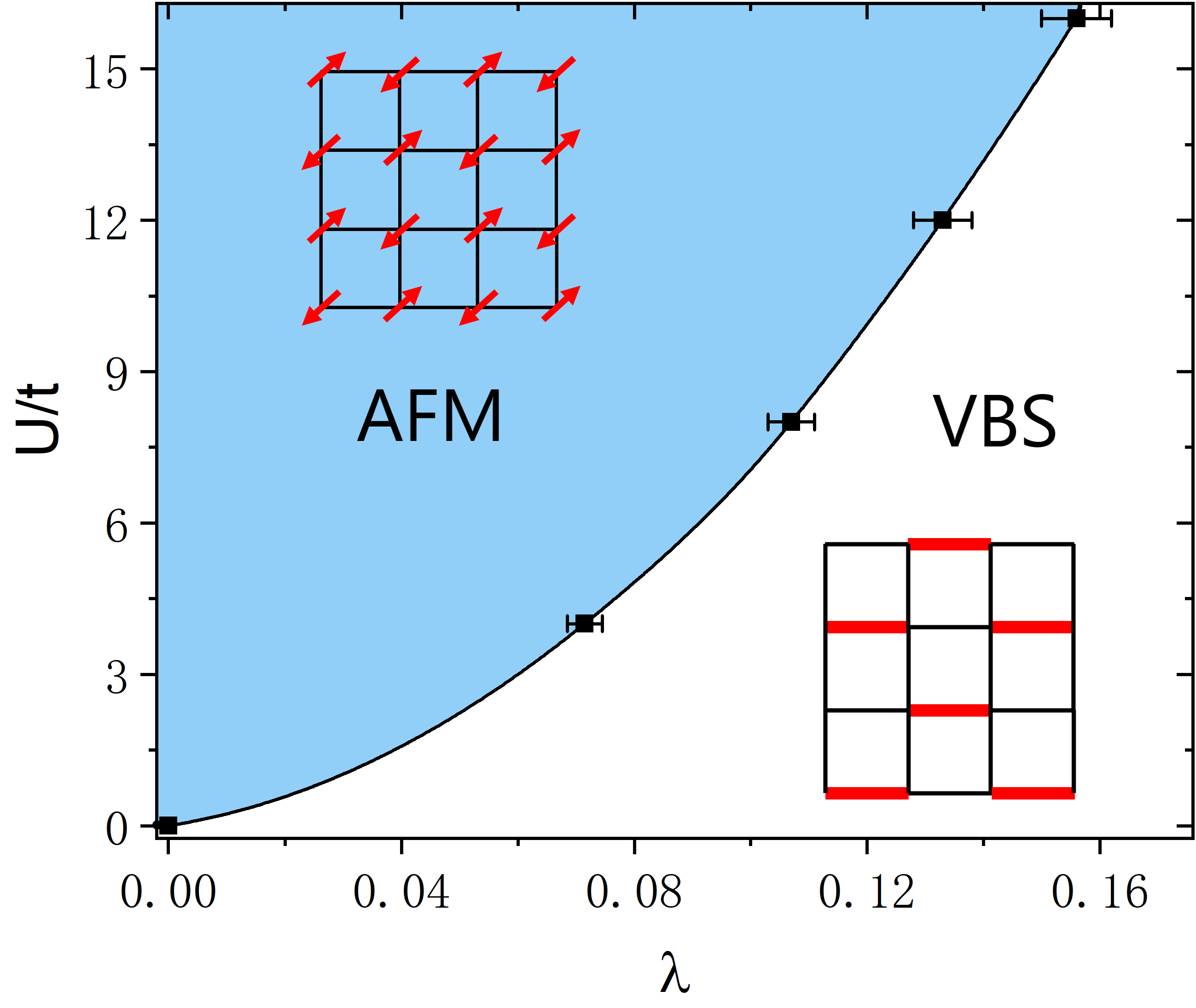}
    \caption{The quantum phase diagram of the half-filled SSHH model in the adiabatic limit, as a function of EPC constant $\lambda$ and Hubbard interaction $U$. The AFM ordering is dominant in the regime where $\lambda$ is small and Hubbard interaction is strong. }
    \label{FigAdiabaticPhaseDiagram}
\end{figure}

Similarly, we compute AFM and VBS correlation ratios for various Hubbard interaction $U$ in the adiabatic limit, and investigate the competition between two ordered phases. We extract the transition value $\lambda_c(U)$ by finite-size scaling analysis of correlation ratio $R_{\mathrm{AFM}}^S,R_{\mathrm{VBS}}^S$, and obtain the ground-state phase diagram of half-filled SSHH model at the adiabatic limit, as the function of $\lambda$ and $U$, as plotted in \Fig{FigAdiabaticPhaseDiagram}. The results clearly demonstrate the interplay between EPC and electronic interaction leads to a dichotomy of the phase diagram, as well as a direct QPT between the two phases breaking spin SU(2) symmetry and lattice $\mathbb{C}_4$ symmetry, respectively. It is also evident that the critical point $\lambda_c$ increase monotonically with $U$, indicating that with growing value of $U$, AFM ordering becomes more stable in the adiabatic limit. Consequently, our numerical simulation shows that, even in the adiabatic limit, AFM ordering is robust for $\lambda<\lambda_c$, where $\lambda_c$ increases monotonically with $U$, as expected.

{\bf Results at finite phonon frequency:} We now move to studying the ground-state properties of SSH-Hubbard model with a generic finite phonon frequency. In the pure SSH model, recent works revealed that the quantum SSH phonons induce AFM long-range order for $\lambda<\lambda_c(\omega)$, with the critical coupling constant $\lambda_c(\omega)$ growing monotonically with $\omega$. It is intriguing to investigate the effects of strong electronic correlation on this SSH phonon induced AFM ordering. In the simulation of finite phonon frequency, we fix Hubbard repulsive at $U=8t$.

\begin{figure}[t]
\subfigure{\label{figAFMratioU8w15}\includegraphics[width= 0.475\linewidth]{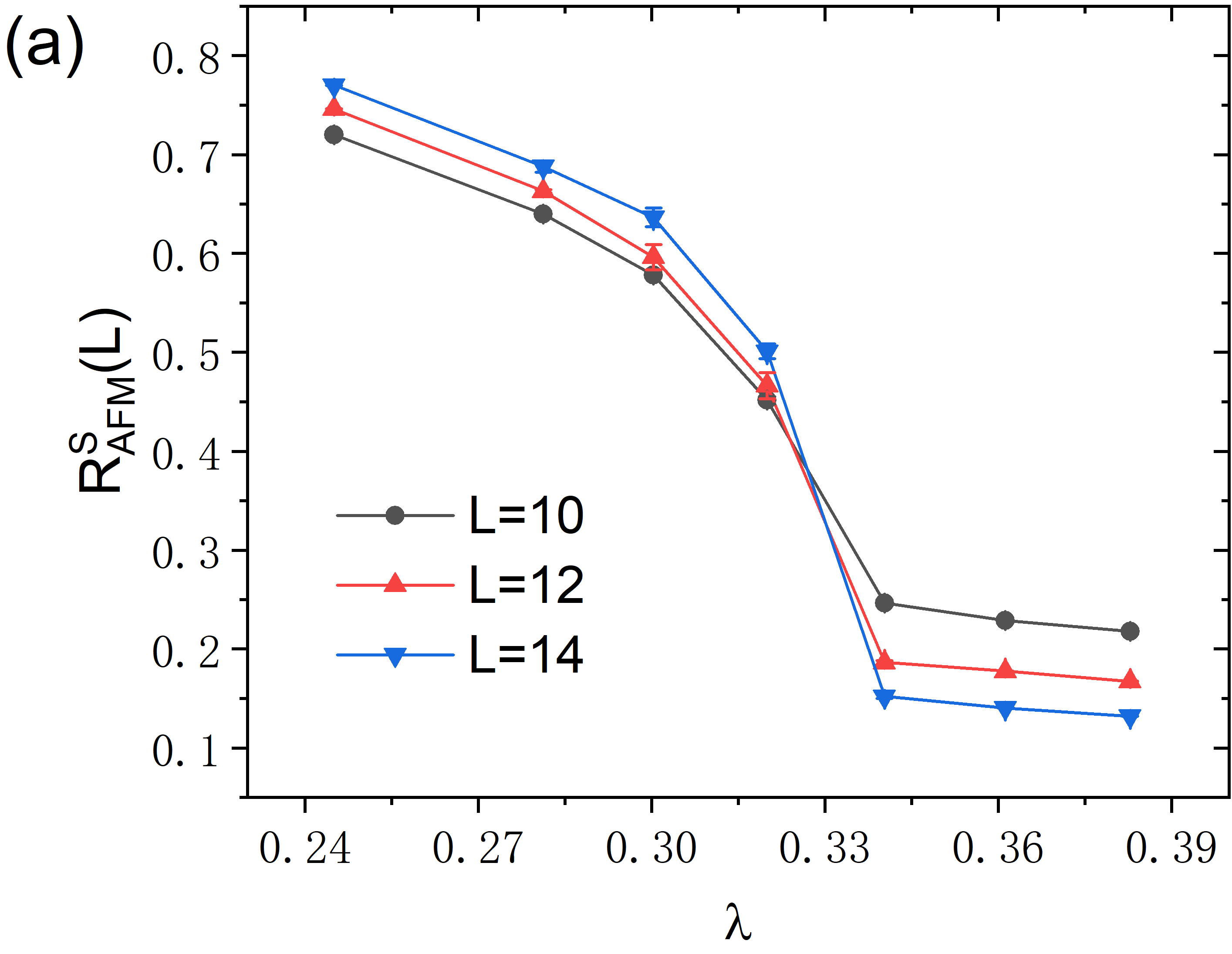}}~~~~
\subfigure{\label{figVBSratioU8w15}\includegraphics[width= 0.475\linewidth]{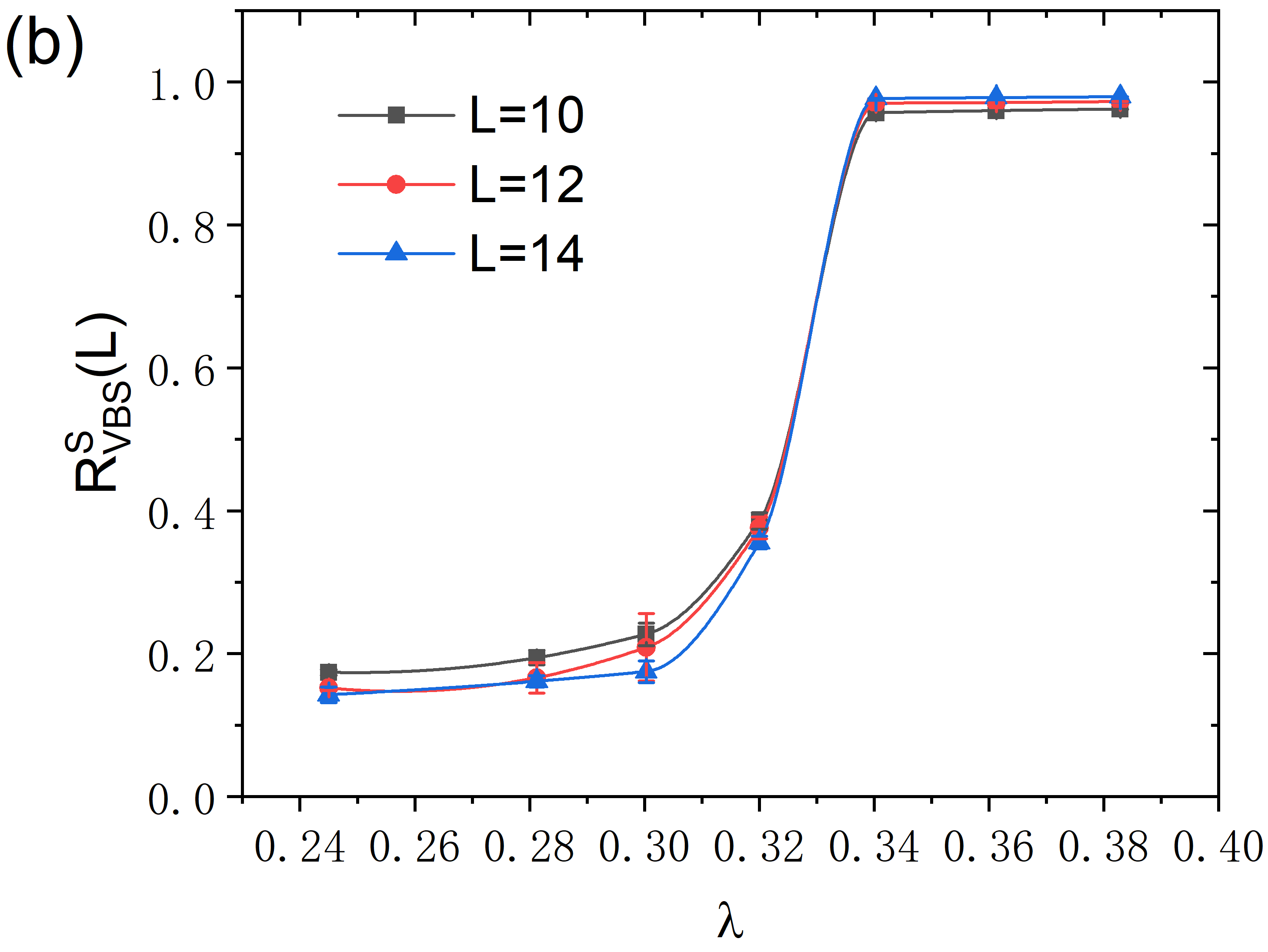}}
\caption{The QMC results of (a) AFM correlation ratio $R_{\mathrm{AFM}}^S$ and (b) VBS correlation ratio $R_{\mathrm{VBS}}^S$, as a function of $\lambda$ for $\omega=1.5$ and $U=8$. Both the crossing points of $R_{\mathrm{AFM}}^S$ and $R_{\mathrm{VBS}}^S$ for different $L$ in (a) and (b) coincide at $\lambda_c\approx 0.33$, indicating a direct phase transition between AFM and VBS ordered phases with increasing EPC constant $\lambda$. }
\label{FigRatioU8w15}
\end{figure}

In \Fig{FigRatioU8w15} we plot the AFM correlation ratio $R_{\mathrm{AFM}}^S$ and VBS correlation ratio $R_{\mathrm{VBS}}^S$ for $\omega=1.5$ and $U=8$. Similar to the pure SSH model, it also suggests AFM ordered phase develops for $\lambda<\lambda_{c,\mathrm{AFM}}$ and VBS order establishes for $\lambda>\lambda_{c,\mathrm{VBS}}$. The crossing of $R_{\mathrm{AFM}}^S$ and $R_{\mathrm{VBS}}^S$ for different system sizes both occur around $\lambda_{c,\mathrm{AFM}}=\lambda_{c,\mathrm{VBS}}\approx 0.33$, indicating a direct QPT between AFM and VBS phases, as is the case in the pure SSH model. However, the critical coupling $\lambda_c$ evidently increases, comparing with $\lambda_c\approx 0.25$ for $\omega = 1.5$ in the pure SSH model \cite{XC2021PRL}. We evaluate the AFM and VBS correlation ratios for various other phonon frequency $\omega$ and extract the corresponding critical EPC constant $\lambda_c(\omega)$ by finite-size scaling. For phonon frequency up to $\omega=1.5$, we confirm that AFM ordering is always enhanced by the presence of strong electronic correlation.

\begin{figure}[t]
\subfigure{\label{figAFMratioU8g14}\includegraphics[width= 0.475\linewidth]{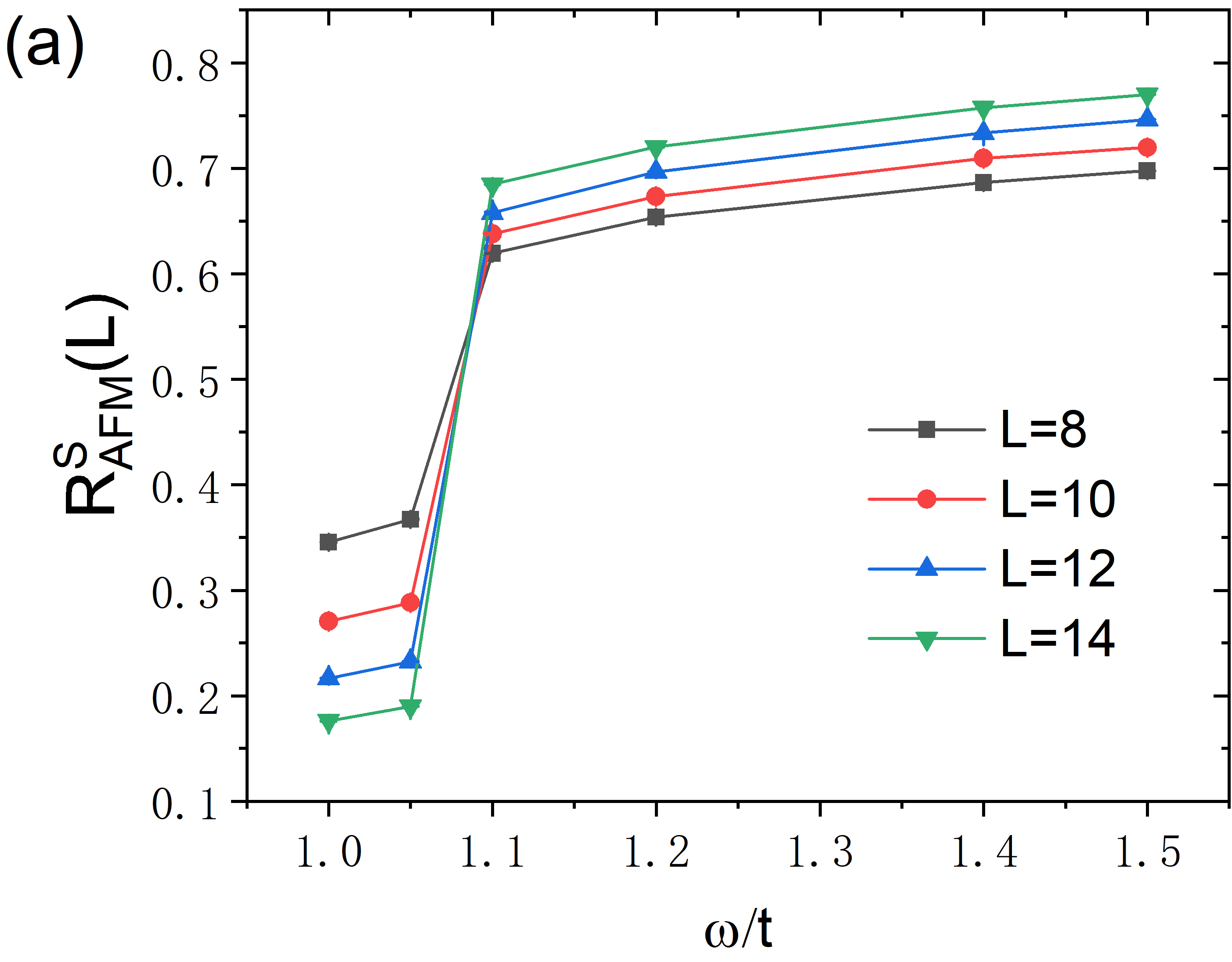}}~~~~
\subfigure{\label{figVBSratioU8g14}\includegraphics[width= 0.475\linewidth]{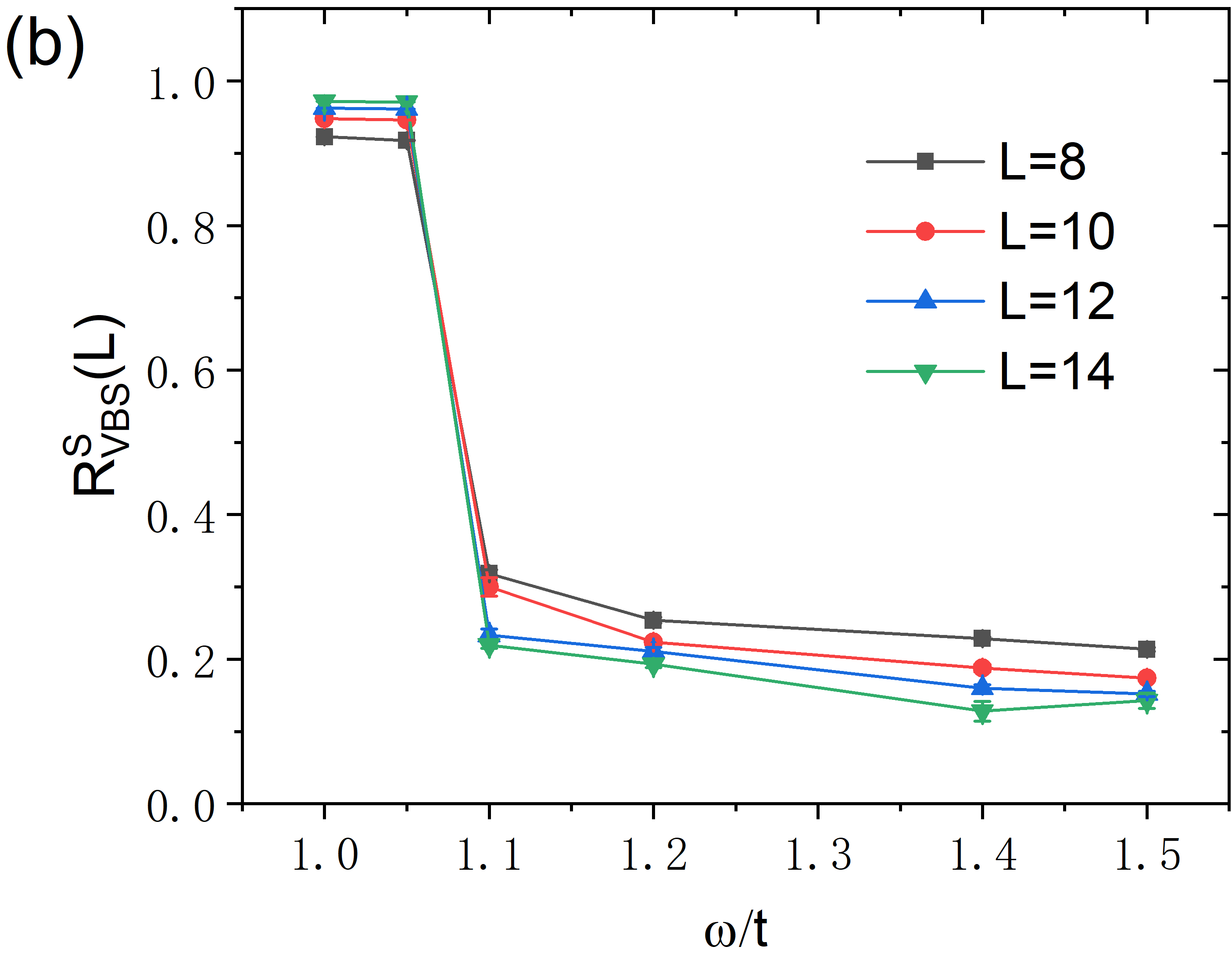}}
\caption{The QMC results of (a) AFM correlation ratio $R_{\mathrm{AFM}}^S$ and (b) VBS correlation ratio $R_{\mathrm{VBS}}^S$, as a function of $\omega$ for $\lambda=0.25$ ($g=1.4$) and $U=8$. Both the crossing points of $R_{\mathrm{AFM}}^S$ and $R_{\mathrm{VBS}}^S$ for different $L$ in (a) and (b) coincide at $\omega_c\approx 1.05$, indicating a direct phase transition between VBS and AFM ordered phases with increasing phonon frequency $\omega$.  }
\label{FigRatioU8g14}
\end{figure}

As discussed above, phonon retardation plays a central role in driving AFM ordering. Here, we investigate the effect of retardation in the presence of electronic correlation by evaluating VBS and AFM correlation ratios with varying phonon frequency but fixed $\lambda$. In \Fig{figAFMratioU8g14} and \Fig{figVBSratioU8g14}, we plot the VBS correlation ratio and AFM correlation ratio for $U=8$ as a function of $\omega$ with fixed $\lambda =0.25$. It is clear that, with increasing phonon frequency, a direct phase transition from VBS to AFM ordered phase at $\omega=\omega_c$. Crucially,  for $\lambda=0.25$ and $U=8$, the crossing points of $R_{\mathrm{AFM}}^\chi$ and $R_{\mathrm{VBS}}^S$ indicates $\omega_c\approx 1.05$, considerably smaller than $\omega_c\approx 1.48$ in the pure SSH model. The results indicates the EPC with bond SSH phonons and onsite electronic interactions can enhance the AFM ordering cooperatively rather than compete with each other.

We investigate the VBS and AFM ordering for various phonon frequency $\omega$ in the presence of strong Hubbard interaction by fixing $U=8$, and obtain the phase boundary between AFM and VBS. Putting the results of finite and zero frequencies together yields the ground-state phase diagram of the SSHH model at half filling for $U=8$, as shown in \Fig{FigPhaseDiagramU8}. Comparing with the phase boundary of the SSH model ($U=0$), the results at $U=8t$ clearly show that AFM ordering is enhanced by the strong electronic correlation, especially in the adiabatic regime where $\omega \ll E_F$. At finite phonon frequency, the SSH phonon driven AFM ordering is robust under the presence of electronic correlation and is actually enhanced by the cooperative interplay between EPC and electronic interactions.

{\bf Conclusions and discussions: }
We systematically explored the ground-state phase diagram of SSHH model in a wide range of phonon frequency, EPC strength, and Hubbard interactions by large-scale projector QMC simulations. 
Our state-of-the-art QMC simulations showed that the phonon-induced AFM ordering is robust in the strongly correlated materials, and furthermore, is enhanced by the strong electronic correlation. Especially, the enhancement of phonon induced AFM order by electronic interaction is more pronounced in the regime where phonon frequency is much smaller compared with electronic Fermi energy, which is practically feasible in most realistic quantum materials. Since increasing experimental progresses reveal that, as mentioned above, EPC could play crucial roles in understanding intriguing physics in strongly correlated materials, including cuprates and iron-based superconductors, the mechanism of cooperative enhancement of AFM ordering by EPC and electronic correlation unveiled in this work could be potentially relevant to AFM insulator in various strongly correlated materials. More importantly, it is intriguing to further investigate cooperative effects of the EPC and electronic correlation on various exotic features arising from doping such AFM insulator phases, such as superconductivity, which is one of promising directions left for future study.

\textit{Acknowledgement}: This work is supported in part by the NSFC under Grant No. 11825404 (X.C. and H.Y.), the MOSTC under Grants No. 2018YFA0305604
and No. 2021YFA1400100 (H.Y.), the CAS
Strategic Priority Research Program under Grant
No. XDB28000000 (H.Y.), and the start-up grant of IOP-CAS (Z.X.L.).

%

\widetext
\section{\large Supplemental Material}
  \setcounter{equation}{0}
  \setcounter{figure}{0}
  \setcounter{table}{0}
  \makeatletter
  \renewcommand{\theequation}{S\arabic{equation}}
  \renewcommand{\thefigure}{S\arabic{figure}}
  \renewcommand{\bibnumfmt}[1]{[S#1]}

\subsection{A. The method of projector quantum Monte Carlo}
In the present paper, we employ projector QMC to study the ground-state properties of the SSHH model described in \Eq{EqOriginalModel} both at finite phonon frequency and in the adiabatic limit (zero frequency). Projector QMC is a numerically exact algorithm designed to detect the ground-state properties of quantum many-body systems. The expectation value of observable $\hat{O}$ in the exact ground state $\ket{\psi_G}$ is evaluated via projecting a trial wave function $\ket{\psi_T}$ along the imaginary-time axis, namely
\beq\label{EqPQMCexpectation}	
\langle\hat{O}\rangle=\frac{\langle\psi_G|\hat{O}|\psi_G\rangle}{\langle\psi_G|\psi_G\rangle} =\lim_{\Theta\ra\infty}\frac{\langle\psi_T|\E{-\Theta H}\hat{O}\E{-\Theta H}|\psi_T\rangle}{\expectation{\psi_T}{\E{-2\Theta H}}{\psi_T}}
\eeq
We should emphasize the algorithm is intrinsically unbiased against the choice of trial wave function $\ket{\psi_T}$ as long as it has a finite overlap with the exact ground state, namely $\braket{\psi_T}{\psi_G} \ne 0$, which is generally satisfied by quantum many-body systems with finite size. In this work we choose $\ket{\psi_T}$ to be the ground state of non-interacting part of the original model \Eq{EqOriginalModel} with only bare electron hopping terms involved.

In practical QMC simulations, the projection length $\Theta$ in \Eq{EqPQMCexpectation} is set to be a finite but sufficiently large value so that the expectation value of each observable in consideration is converged against increasing $\Theta$ . In this work we set $\Theta=36/t, 40/t, 42/t, 46/t, 50/t$ for system size $L=6,8,10,12,14$ accordingly, each of which has been checked to be large enough for convergence. Similar to finite temperature algorithm, Trotter decomposition is implemented here by discretizing $\Theta$ into small imaginary time spacing $\Delta\tau=\Theta/L_\tau$ at the price of introducing a Trotter discretization error scaling as $\Delta\tau^2$. In this work we set $\Delta\tau=0.1/t$. The convergence of discretization has also been checked by comparing the results to smaller $\Delta\tau$, e.g. $\Delta\tau=0.05$.

In model \Eq{EqOriginalModel} both EPC and electronic interaction are involved. The coupling term between electrons and SSH phonons is quadratic in fermionic operators so that we can compute electron's Green's function straightforwardly under each phonon configuration. On the other hand, for Hubbard interaction one has to perform Hubbard-Stratonovich (HS) transformation to decompose the interaction term into fermion bilinears at the price of introducing an auxiliary boson field. We implement a SU(2) symmetric H-S transformation
\beq\label{EqHStransformation}
\E{-\Delta\tau U\inc{n_{i\uparrow}-\frac{1}{2}}\inc{n_{i\downarrow}-\frac{1}{2}}}=\gamma\sum_{s=\pm 1}\E{\imth\alpha s\inc{n_{i\uparrow}+n_{i\downarrow}-1}}
\eeq
on each lattice site\cite{Assaadnote}, where $\gamma=\frac{1}{2}\E{\Delta\tau U/4}$ and $\cos{\alpha}=\E{-\Delta\tau U/2}$ are constants, and $s= \pm 1$ serves as discretized auxiliary field. In this work we sample both phonon and auxiliary field configurations by Metropolis algorithm. For generic finite phonon frequency, we need to sample the phonon configuration on a space-time lattice. In the adiabatic limit, the kinetic term of phonon vanishes, namely phonon fields are static, allowing us to sample the phonon configurations only depending on spatial coordinates.

\begin{figure}[t]
\begin{minipage}{0.3\linewidth}
\centering
\subfigure{\label{figAFMratioU8w10}\includegraphics[width=5.5cm]{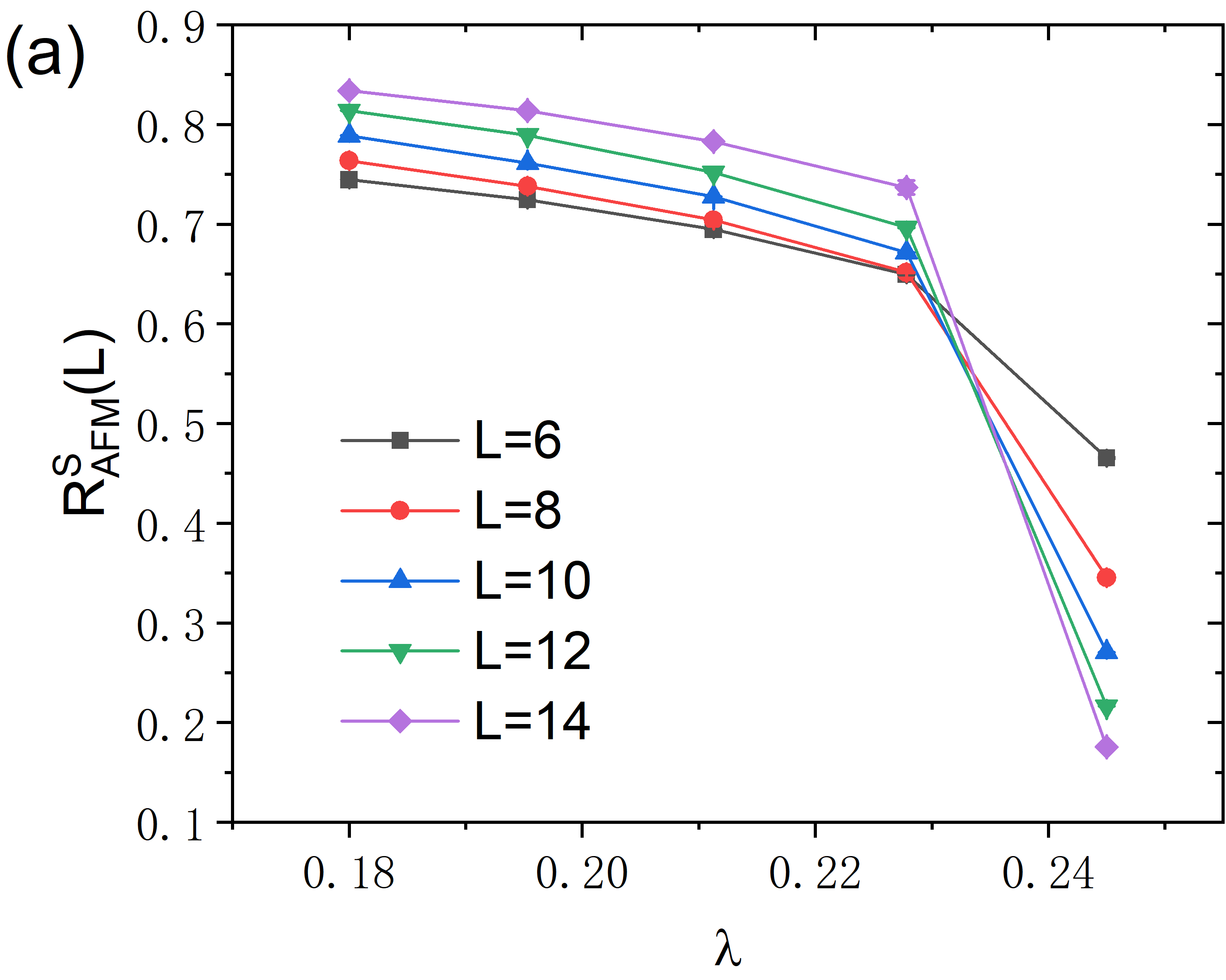}}\\
\subfigure{\label{figVBSratioU8w10}\includegraphics[width=5.5cm]{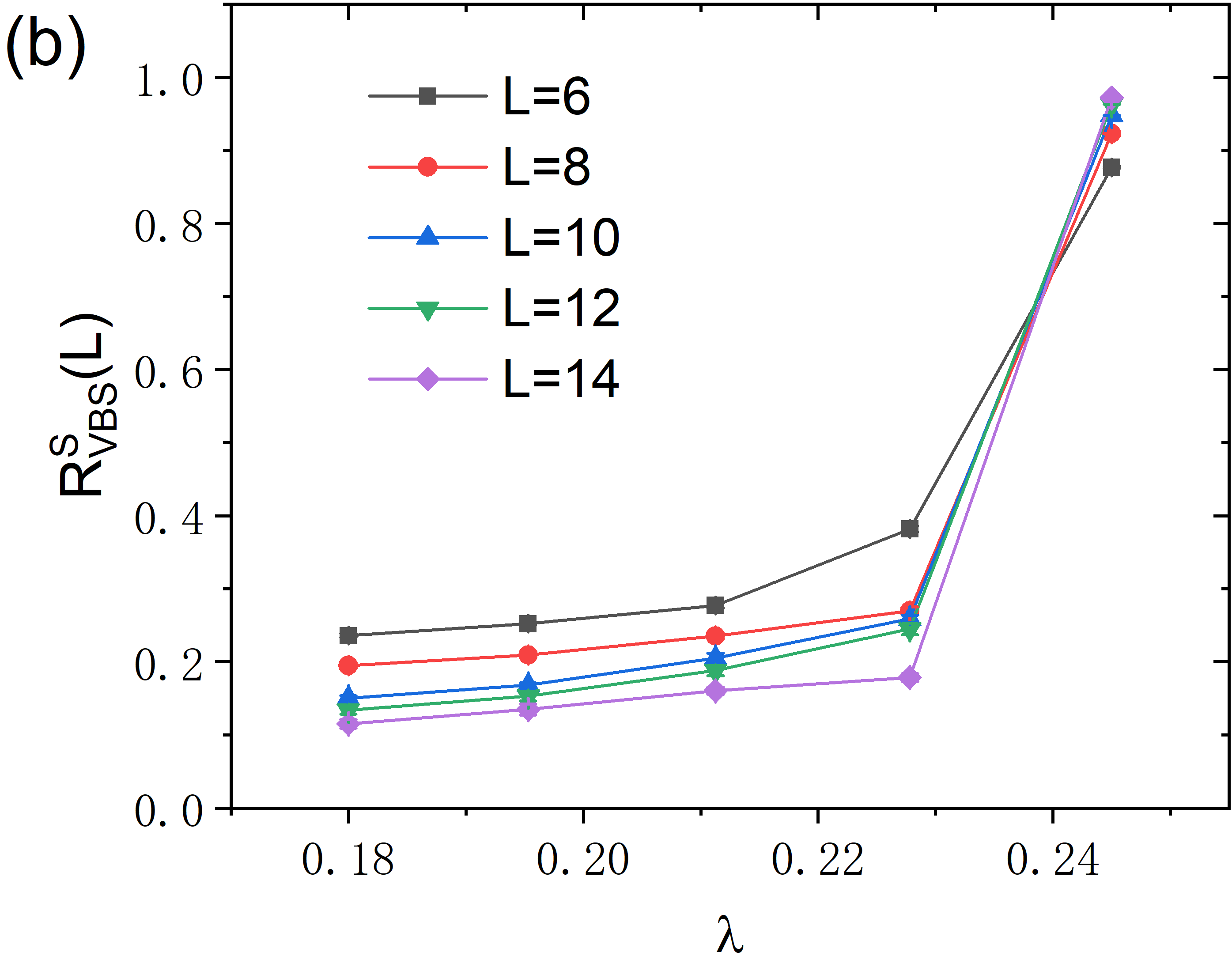}}
\end{minipage}~~~~~~~~
\hspace{0.008 \linewidth}
\begin{minipage}{0.3\linewidth}
\centering
\subfigure{\label{figAFMratioU8w07}\includegraphics[width=5.5cm]{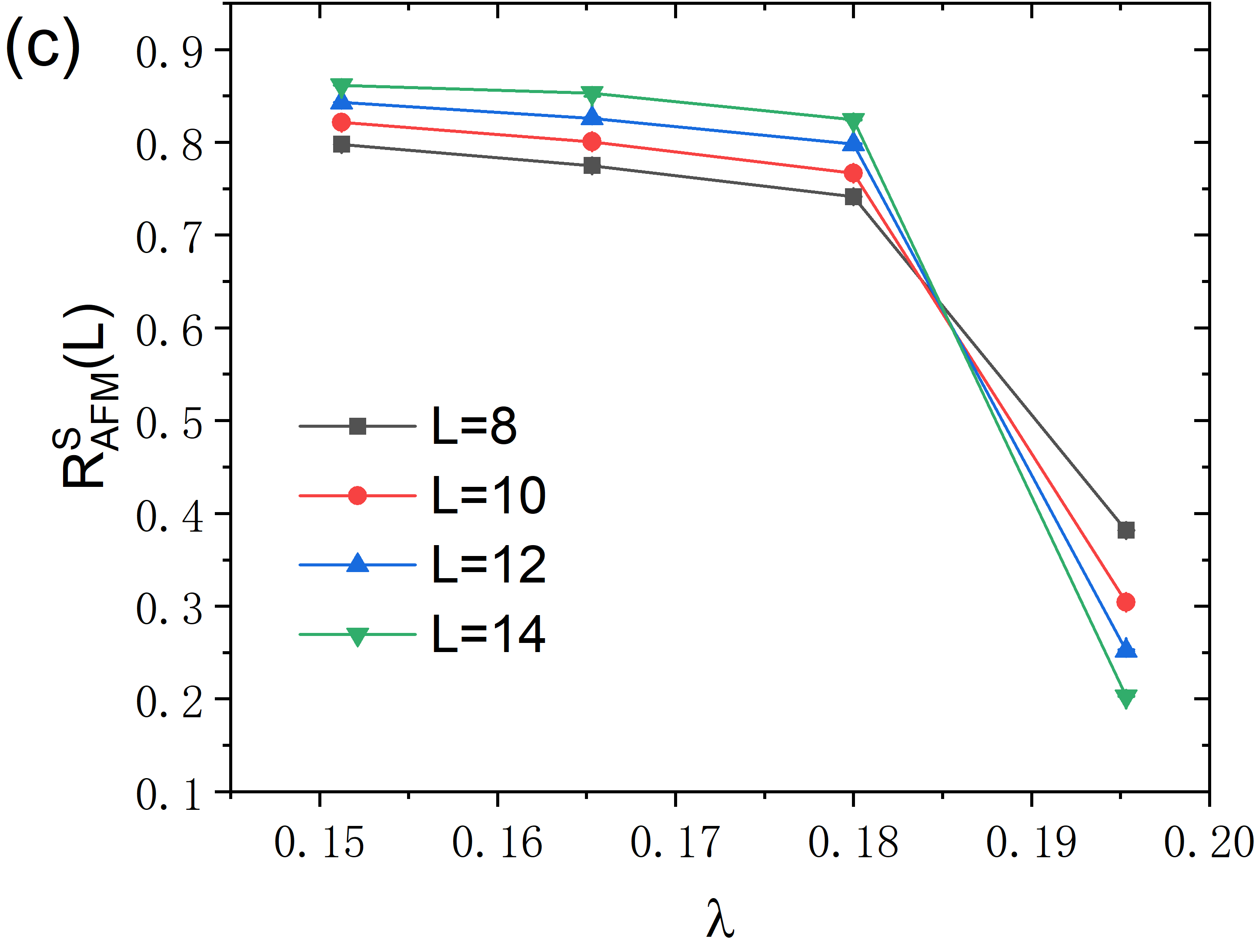}}\\
\subfigure{\label{figVBSratioU8w07}\includegraphics[width=5.5cm]{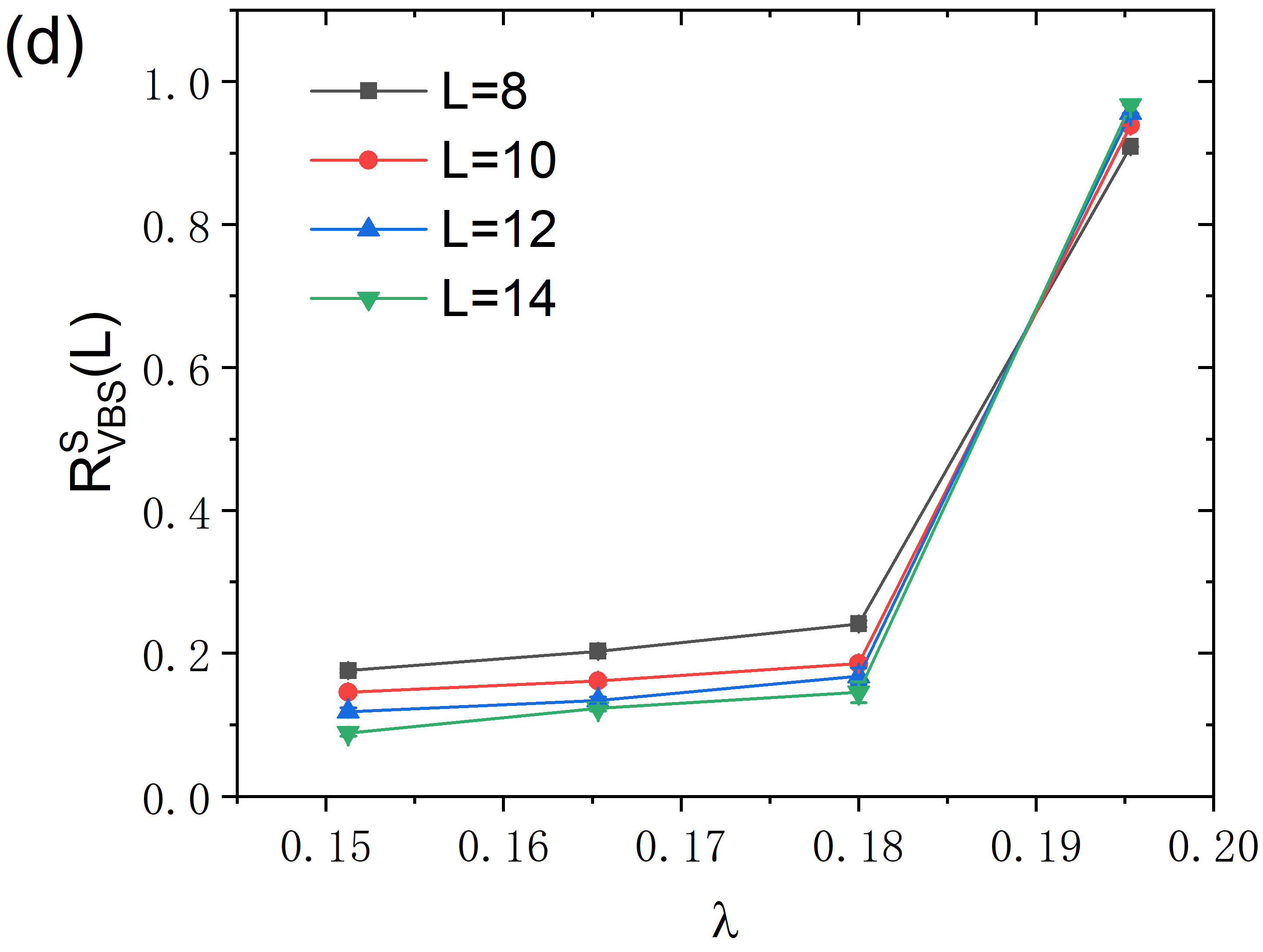}}
\end{minipage}~~~~~~~~
\caption{The QMC results of correlation ratio as a function $\lambda$ for AFM in (a) and (c) and for VBS in (b) and (d). Phonon frequency is fixed at $\omega=1.0t$ in (a) and (b) and $\omega=0.7t$ in (c) and (d). The crossing point in each figure coincide with the phase boundary shown in \Fig{FigPhaseDiagramU8} in the main text. }
\label{figSMCorrRatio}
\end{figure}

\subsection{B. The definition of order parameter}
The AFM and VBS order parameters for finite size $N=L^2$ evaluated in the main text are given by
\begin{align}
\hat{O}_{\mathrm{AFM}}\inc{\v{q}}&=\frac{1}{L^2}\sum_j \E{\imth \v{q}\cdot \v{R}_j} \hat{S}_j^z \label{EqAFMorder}\\
\hat{O}_{\mathrm{VBS}}\inc{\v{q}}&=\frac{1}{L^2}\sum_j \E{\imth \v{q}\cdot \v{R}_j} \inc{\hat{B}_{j,\hat{x}}+\imth \hat{B}_{j,\hat{y}}} \label{EqVBSorder}
\end{align}
where $\hat{B}_{j,\delta}= \sum_\sigma c^\dagger_{j,\s}c_{j+\delta,\s}+h.c.$ is the kinetic operator on $\delta=\hat{x}, \hat{y}$ bonds. The staggered VBS order breaks lattice $\mathbb{C}_4$ symmetry while AFM breaks spin SU(2) rotational symmetry. For both VBS and AFM order on the square lattice, the ordering wave vector is $\v{Q}=\inc{\pi,\pi}$. The corresponding structure factor for each order parameter is defined as $S(\v{q})=\langle |\hat{O}(\v{q})|^2\rangle$, which is peaked at $\v{Q}=\inc{\pi,\pi}$.

\begin{figure}[t]
\begin{minipage}{0.3\linewidth}
\centering
\subfigure{\label{figAFMsusratioU8w15}\includegraphics[width=5.5cm]{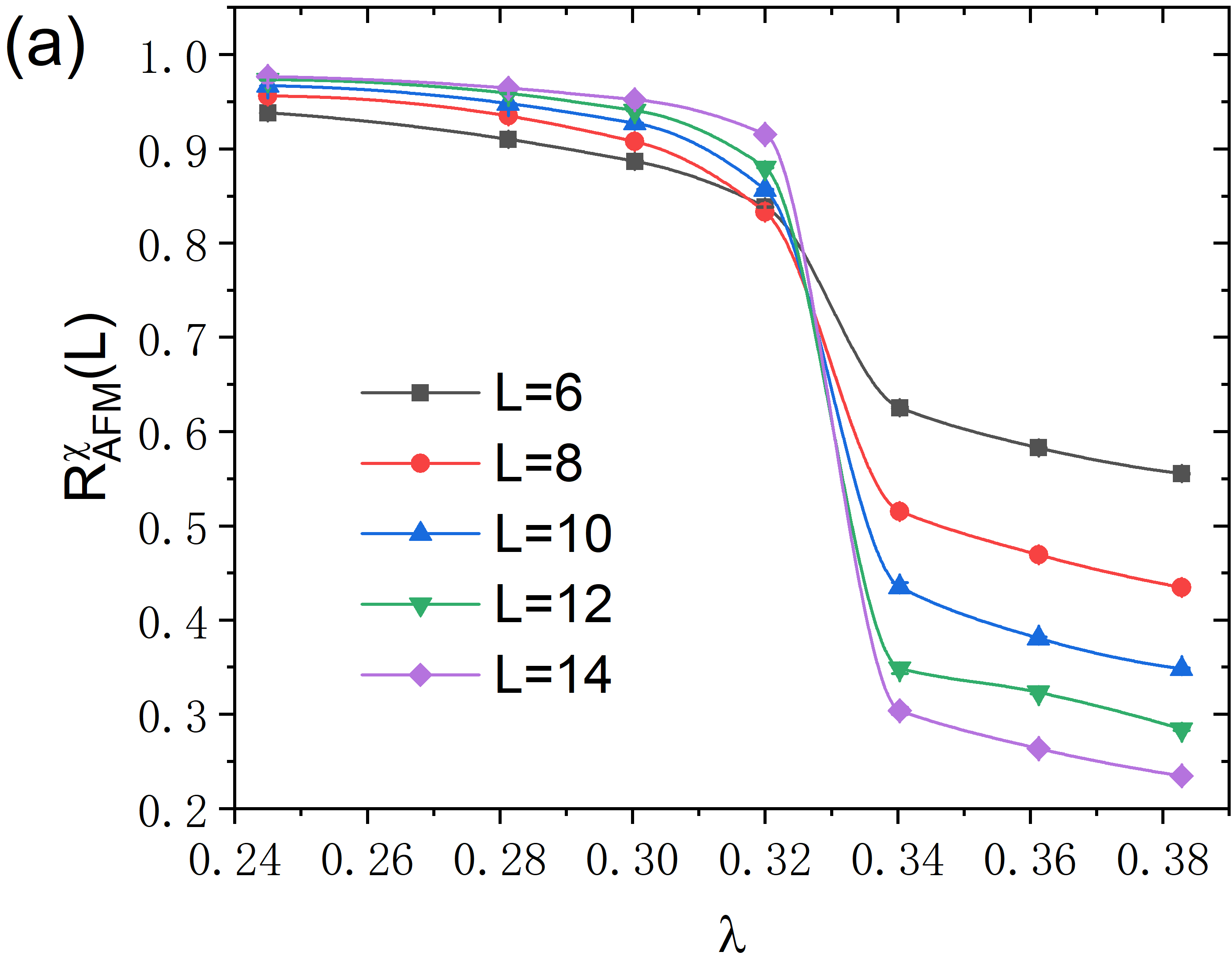}}\\
\subfigure{\label{figAFMsusratioU8w10}\includegraphics[width=5.5cm]{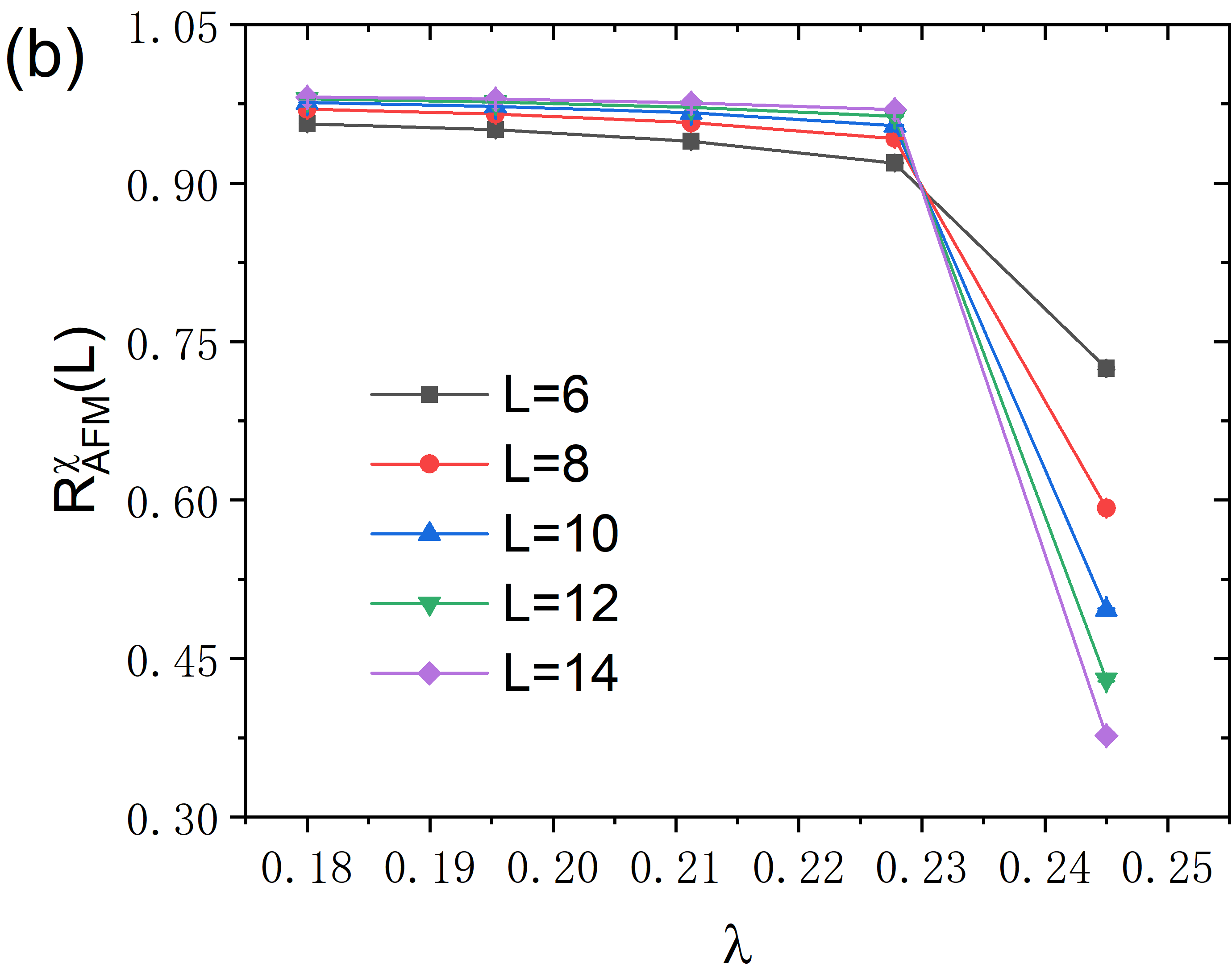}}
\end{minipage}~~~~~~~~
\hspace{0.008 \linewidth}
\begin{minipage}{0.3\linewidth}
\centering
\subfigure{\label{figAFMsusratioU8w07}\includegraphics[width=5.5cm]{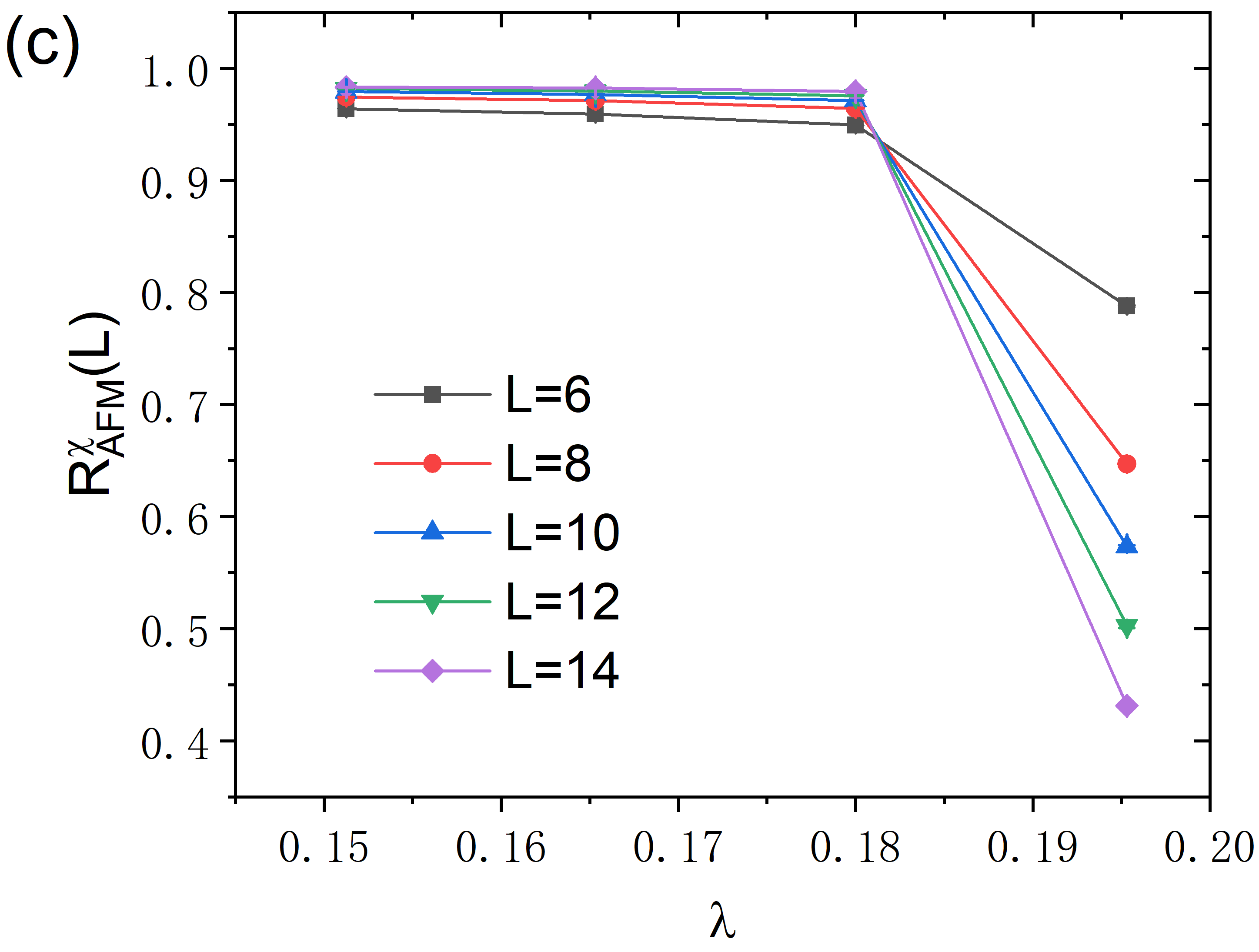}}\\
\subfigure{\label{figAFMsusratioU8g14}\includegraphics[width=5.5cm]{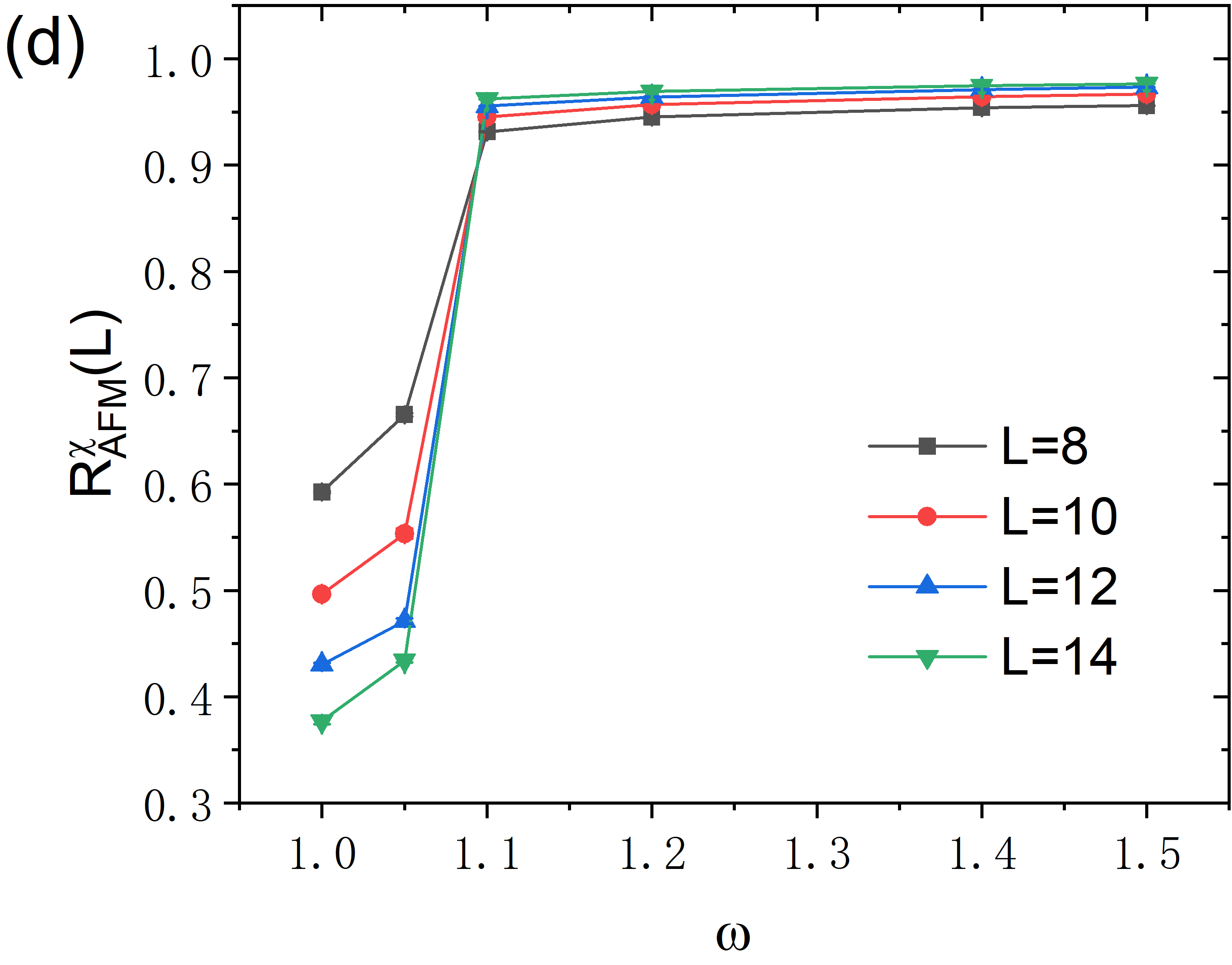}}
\end{minipage}~~~~~~~~
\caption{The QMC results of AFM susceptibility ratio as a function of coupling $\lambda$ in (a), (b) and (c), and as a function of phonon frequency $\omega$ in (d). Phonon frequencies are fixed at (a) $\omega=1.5t$, (b) $\omega=1.0t$ and (c) $\omega=0.7t$. In (d) the coupling is fixed at $\lambda=0.25$. }
\label{figSMSusRatio}
\end{figure}

\subsection{C. QMC results for different phonon frequencies and spin susceptibility ratio }
In the main text we present the results of AFM and VBS correlation ratios at fixed phonon frequency at $\omega=0$ and $\omega=1.5t$. Here, we provide QMC results for other phonon frequencies. In \Fig{figSMCorrRatio}, we plot the AFM and VBS correlation ratios for $\omega=1.0t$ and $\omega=0.7t$. As discussed in the main text, the crossing point for different system sizes $L$ indicates the transition point. From \Fig{figSMCorrRatio} we obtain the critical coupling of AFM-VBS transition $\lambda_c\approx 0.19$ and $\lambda_c\approx 0.23$ for $\omega=0.7t$ and $\omega=1.0t$, respectively, both of which are depicted in \Fig{FigPhaseDiagramU8} in the main text.

We also compute the spin susceptibility which may have smaller finite size corrections than equal time correlations, as mentioned in the main text. For projector QMC algorithm, the spin susceptibility can be computed as an integration over the imaginary-time interval $\tau_M$
\beq\label{EqSusceptibilityDef}
\chi_{\mathrm{AFM}}\inc{\v{q}}= \int_{\Theta-\tau_M/2}^{\Theta+\tau_M/2} \dif\tau\,\langle\hat{O}_{\mathrm{AFM}} \inc{\v{q},\tau}\hat{O}_{\mathrm{AFM}}\inc{\v{q},0}\rangle.
\eeq
In projector QMC algorithm, the imaginary-time interval $\tau_M$ should be large enough to guarantee the convergence of $\chi$, while the correlators $\langle \hat{O}\inc{\v{q},\tau}\hat{O}\inc{\v{q},0}\rangle$ should also be computed after sufficiently long projection. Thus, in practice we choose $\tau_M\approx\Theta/4$ and integrate over the correlators from $\Theta-\tau_M/2$ to $\Theta+\tau_M/2$. In \Fig{figSMSusRatio} we show the AFM susceptibility ratio for various phonon frequencies or coupling strength. It is evident that the critical coupling $\lambda_c$ and critical frequency obtained from \Fig{figSMSusRatio} are consistent with the results shown in \Fig{FigPhaseDiagramU8}.
\end{document}